\documentclass{aastex631}

\begin{document}

\title{Photometry of Type II Supernova SN 2023ixf with a Worldwide Citizen Science Network}

\author[0000-0001-6629-5399]{Lauren A. Sgro}
\affiliation{SETI Institute, 339 N Bernardo Ave Suite 200, Mountain View, CA 94043, USA}
\affiliation{Unistellar, 5 all\'ee Marcel Leclerc, b\^{a}timent B, Marseille, 13008, France}

\author[0000-0002-0792-3719]{Thomas M. Esposito}
\affiliation{SETI Institute, 339 N Bernardo Ave Suite 200, Mountain View, CA 94043, USA}
\affiliation{Unistellar, 5 all\'ee Marcel Leclerc, b\^{a}timent B, Marseille, 13008, France}
\affiliation{Department of Astronomy, University of California, Berkeley, CA 94720, USA}

\author[0000-0002-0973-4276]{Guillaume Blaclard}
\affiliation{Unistellar, 5 all\'ee Marcel Leclerc, b\^{a}timent B, Marseille, 13008, France}

\author[0000-0001-6395-6702]{Sebastian Gomez}
\affiliation{Space Telescope Science Institute, 3700 San Martin Drive, Baltimore, MD 21218, USA}

\author[0000-0001-7016-7277]{Franck Marchis}
\affiliation{SETI Institute, 339 N Bernardo Ave Suite 200, Mountain View, CA 94043, USA}
\affiliation{Unistellar, 5 all\'ee Marcel Leclerc, b\^{a}timent B, Marseille, 13008, France}

\author[0000-0003-3460-0103]{Alexei V. Filippenko}
\affiliation{Department of Astronomy, University of California, Berkeley, CA 94720, USA}

\author[0000-0002-9427-0014]{Daniel O’Conner Peluso}
\affiliation{SETI Institute, 339 N Bernardo Ave Suite 200, Mountain View, CA 94043, USA}
\affiliation{Centre for Astrophysics, University of Southern Queensland, Toowoomba, QLD, Australia}

\author[0000-0002-7491-7052]{Stephen S. Lawrence}
\affiliation{Dept. of Physics \& Astronomy, 151 Hofstra University, Hempstead, NY 11549, USA}
\affiliation{Unistellar Citizen Scientist}

\author{Aad Verveen} 
\affiliation{Unistellar Citizen Scientist}
\author{Andreas Wagner}
\affiliation{Unistellar Citizen Scientist}
\author{Anouchka Nardi}
\affiliation{Unistellar Citizen Scientist}
\author{Barbara Wiart}
\affiliation{Unistellar Citizen Scientist}
\author{Benjamin Mirwald}
\affiliation{Unistellar Citizen Scientist}
\author{Bill Christensen}
\affiliation{Unistellar Citizen Scientist}
\author{Bob Eramia}
\affiliation{Unistellar Citizen Scientist}
\author[0000-0002-9578-5765]{Bruce Parker} 
\affiliation{Unistellar Citizen Scientist}
\author[0000-0003-4091-0247]{Bruno Guillet} 
\affiliation{Unistellar Citizen Scientist}
\author{Byungki Kim}
\affiliation{Unistellar Citizen Scientist}
\author[0009-0003-4751-4906]{Chelsey A. Logan}
\affiliation{Unistellar Citizen Scientist}
\author[0000-0001-7014-1843]{Christopher C. M. Kyba}
\affiliation{Unistellar Citizen Scientist}
\author{Christopher Toulmin}
\affiliation{Unistellar Citizen Scientist}
\author{Claudio G. Vantaggiato}
\affiliation{Unistellar Citizen Scientist}
\author{Dana Adhis}
\affiliation{Unistellar Citizen Scientist}
\author{Dave Gary}
\affiliation{Unistellar Citizen Scientist}
\author{Dave Goodey}
\affiliation{Unistellar Citizen Scientist}
\author{David Dickinson}
\affiliation{Unistellar Citizen Scientist}
\author{David Koster}
\affiliation{Unistellar Citizen Scientist}
\author{Davy Martin}
\affiliation{Unistellar Citizen Scientist}
\author[0009-0003-4759-1434]{Eliud Bonilla}
\affiliation{Unistellar Citizen Scientist}
\author{Enner Chung}
\affiliation{Unistellar Citizen Scientist}
\author{Eric Miny}
\affiliation{Unistellar Citizen Scientist}
\author[0000-0002-6818-6599]{Fabrice Mortecrette}
\affiliation{Unistellar Citizen Scientist}
\author[0009-0003-6483-0433]{Fadi Saibi}
\affiliation{Unistellar Citizen Scientist}
\author[0009-0008-5454-6929]{Francois O. Gagnon}
\affiliation{Unistellar Citizen Scientist}
\author{François Simard}
\affiliation{Unistellar Citizen Scientist}
\author{Gary Vacon}
\affiliation{Unistellar Citizen Scientist}
\author{Georges Simard}
\affiliation{Unistellar Citizen Scientist}
\author{Gerrit Dreise}
\affiliation{Unistellar Citizen Scientist}
\author{Hiromi Funakoshi}
\affiliation{Unistellar Citizen Scientist}
\author{Janet Vacon}
\affiliation{Unistellar Citizen Scientist}
\author{James Yaniz}
\affiliation{Unistellar Citizen Scientist}
\author[0009-0000-0969-2216]{Jean-Charles Le Tarnec}
\affiliation{Unistellar Citizen Scientist}
\author[0000-0002-1908-6057]{Jean-Marie Laugier}
\affiliation{Unistellar Citizen Scientist}
\author{Jennifer L. W. Siders}
\affiliation{Unistellar Citizen Scientist}
\author{Jim Sweitzer}
\affiliation{Unistellar Citizen Scientist}
\author{Jiri Dvoracek}
\affiliation{Unistellar Citizen Scientist}
\author[0009-0008-1227-5083]{John Archer}
\affiliation{Unistellar Citizen Scientist}
\author{John Deitz}
\affiliation{Unistellar Citizen Scientist}
\author{John K. Bradley}
\affiliation{Unistellar Citizen Scientist}
\author[0000-0002-9297-5133]{Keiichi Fukui}
\affiliation{Unistellar Citizen Scientist}
\author[0000-0003-0690-5508]{Kendra Sibbernsen}
\affiliation{Unistellar Citizen Scientist}
\author{Kevin Borrot}
\affiliation{Unistellar Citizen Scientist}
\author{Kevin Cross}
\affiliation{Unistellar Citizen Scientist}
\author[0009-0004-1079-7196]{Kevin Heider}
\affiliation{Unistellar Citizen Scientist}
\author{Koichi Yamaguchi}
\affiliation{Unistellar Citizen Scientist}
\author[0000-0001-8058-7443]{Lea A. Hirsch}
\affiliation{Unistellar Citizen Scientist}
\author[0000-0003-3046-9187]{Liouba Leroux}
\affiliation{Unistellar Citizen Scientist}
\author[0000-0002-3278-9590]{Mario Billiani}
\affiliation{Unistellar Citizen Scientist}
\author{Markus Lorber}
\affiliation{Unistellar Citizen Scientist}
\author{Martin J. Smallen}
\affiliation{Unistellar Citizen Scientist}
\author{Masao Shimizu}
\affiliation{Unistellar Citizen Scientist}
\author{Masayoshi Nishimura}
\affiliation{Unistellar Citizen Scientist}
\author[0000-0001-8337-0020]{Matthew Ryno}
\affiliation{Unistellar Citizen Scientist}
\author{Michael Cunningham}
\affiliation{Unistellar Citizen Scientist}
\author{Michael Gagnon}
\affiliation{Unistellar Citizen Scientist}
\author[0000-0003-3462-7533]{Michael Primm}
\affiliation{Unistellar Citizen Scientist}
\author{Michael Rushton}
\affiliation{Unistellar Citizen Scientist}
\author[0000-0001-7412-0193]{Michael Sibbernsen}
\affiliation{Unistellar Citizen Scientist}
\author{Mike Mitchell}
\affiliation{Unistellar Citizen Scientist}
\author{Neil Yoblonsky}
\affiliation{Unistellar Citizen Scientist}
\author{Niniane Leroux}
\affiliation{Unistellar Citizen Scientist}
\author{Olivier Clerget}
\affiliation{Unistellar Citizen Scientist}
\author[0000-0002-9418-3754]{Ozren Stojanović}
\affiliation{Unistellar Citizen Scientist}
\author{Patrice Unique}
\affiliation{Unistellar Citizen Scientist}
\author[0000-0003-1371-4232]{Patrick Huth}
\affiliation{Unistellar Citizen Scientist}
\author[0000-0002-8703-6430]{Raymund John Ang}
\affiliation{Unistellar Citizen Scientist}
\author{Regis Santoni}
\affiliation{Unistellar Citizen Scientist}
\author{Robert Foster}
\affiliation{Unistellar Citizen Scientist}
\author{Roberto Poggiali}
\affiliation{Unistellar Citizen Scientist}
\author[0009-0004-0672-2255]{Ruyi Xu}
\affiliation{Unistellar Citizen Scientist}
\author[0000-0001-7029-644X]{Ryuichi Kukita}
\affiliation{Unistellar Citizen Scientist}
\author{Sanja Šćepanović}
\affiliation{Unistellar Citizen Scientist}
\author[0009-0006-8494-5408]{Sophie Saibi}
\affiliation{Unistellar Citizen Scientist}
\author[0000-0003-0404-6279]{Stefan Will}
\affiliation{Unistellar Citizen Scientist}
\author{Stephan Latour}
\affiliation{Unistellar Citizen Scientist}
\author{Stephen Haythornthwaite}
\affiliation{Unistellar Citizen Scientist}
\author{Sylvain Cadieux}
\affiliation{Unistellar Citizen Scientist}
\author[0000-0002-9089-6853]{Thoralf Müller}
\affiliation{Unistellar Citizen Scientist}
\author[0000-0003-1368-861X]{Tze Yang Chung}
\affiliation{Unistellar Citizen Scientist}
\author{Yoshiya Watanabe}
\affiliation{Unistellar Citizen Scientist}
\author{Yvan Arnaud}
\affiliation{Unistellar Citizen Scientist}

\begin{abstract}

We present highly sampled photometry of the supernova (SN) 2023ixf, a Type II SN in M101, beginning 2~days before its first known detection. To gather these data, we enlisted the global Unistellar Network of citizen scientists. These 252 observations from 115 telescopes show the SN's rising brightness associated with shock emergence followed by gradual decay. We measure a peak $M_{V} = -18.18 \pm 0.09$~mag at 2023-05-25 21:37 UTC in agreement with previously published analyses.

\end{abstract}


\section{Introduction} \label{sec:intro}
Type II supernovae (SNe) are hydrogen-rich, core-collapse SNe (see \citealt{filippenko1997optical} for a review of SN classification) and are among the most commonly observed SNe (e.g., \citealt{li2011nearby}). Despite their prevalence, early-time observations of these SNe are rarely available owing to the cadence of large surveys and other factors. Nevertheless, data within days after shock breakout, in which the shock wave from the collapsing core reaches the stellar photosphere, are imperative for gaining an understanding of the progenitor and explosion physics \citep{waxman2017handbook}.

\citet{2023TNSTR1158....1I} reported discovery of a SN in M101 (redshift $z=0.0008$) in observations from 2023-05-19 17:27 (UTC dates are used throughout this paper). This SN, designated 2023ixf, offered an opportunity for amateur and professional astronomers to collect data promptly. The earliest known detection was found during spontaneous observations by \citet{{2023TNSAN.130....1M}} at 2023-05-18 20:29. See \citet{2023arXiv230606097H} and references therein for a review of early-time photometry. 

\citet{2023TNSAN.119....1P} revealed SN 2023ixf as a Type II SN. The progenitor candidate has been identified in archival {\it Spitzer Space Telescope} and {\it Hubble Space Telescope} images \citep{szalai2023spitzer,soraisam,pledger} as a luminous red supergiant with a dense shell of circumstellar material and long-period variability in near-to-mid-infrared wavelengths \citep{Jacobson-Galan2023,Jencson,Kilpatrick}. Future studies will further constrain the progenitor and SN to increase understanding of Type II SNe.

\section{Observations \& Data Reduction} \label{sec:obs}

All data used in this work were taken by the Unistellar Network, comprised of observers worldwide who use Unistellar telescopes \textemdash~ 11.4~cm aperture digital, smart telescopes \citep{marchis2020unistellar}. Each telescope employs a CMOS sensor sensitive to blue, green, and red bandpasses via a Bayer filter. Uniform optical properties simplify combination of data from multiple telescopes, making possible results such as those described by \citet{graykowski2023light}, \citet{perrocheau202216}, and \citet{peluso2023unistellar}. 

\begin{figure}[ht!]
\plotone{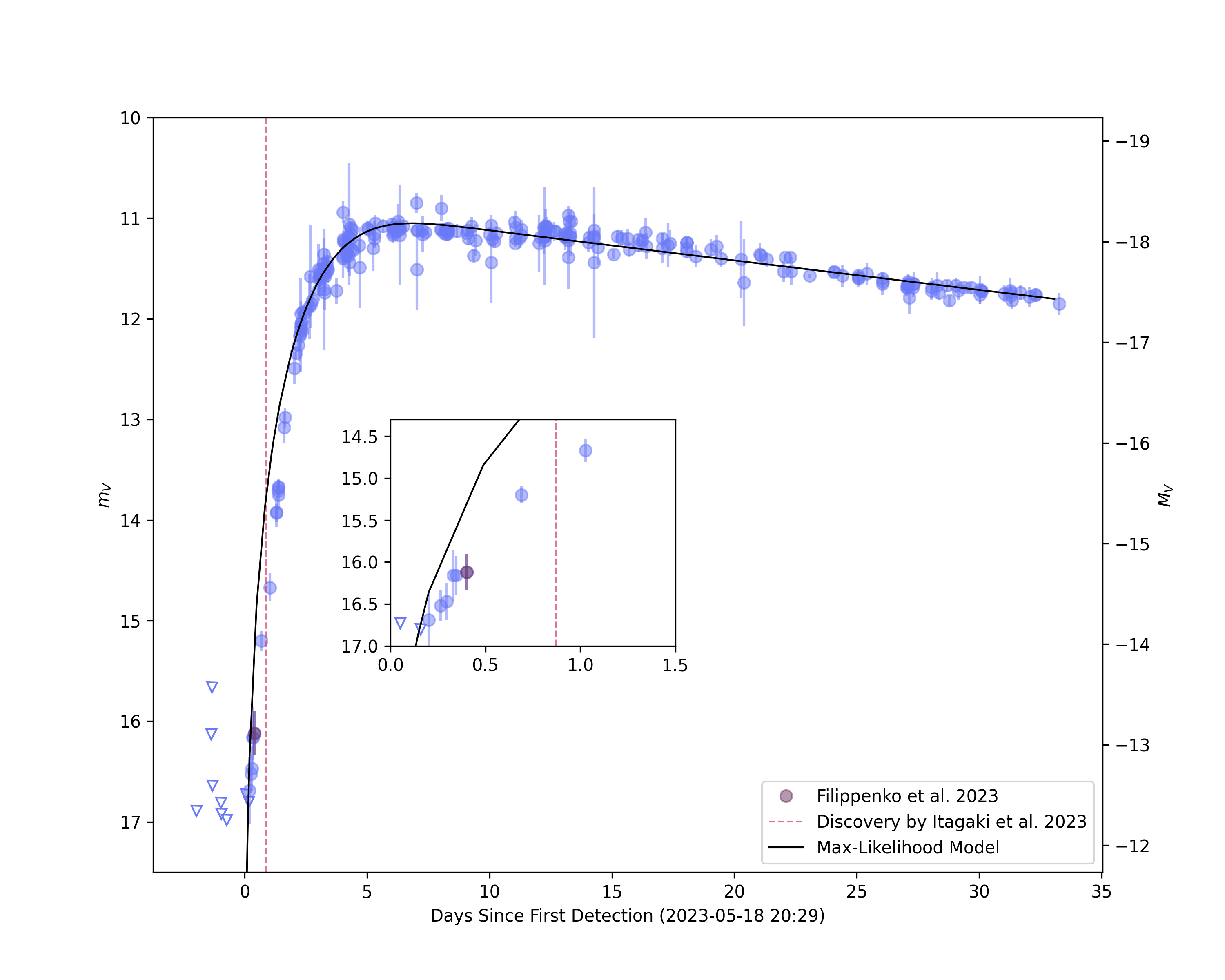}
\caption{Light curve of SN 2023ixf. Circles represent detections and triangles represent $3\sigma$ lower magnitude limits. Inset highlights the first 1.5~days after earliest detection.
\label{fig:LC}}
\end{figure}

Unistellar telescopes can record data in two modes, Enhanced Vision (EV) and Science mode, which were utilized to measure the light curve of SN 2023ixf presented in Figure \ref{fig:LC}. The EV data were taken during prediscovery and postdiscovery serendipitous imaging of M101, while coordinated Science observations commenced ${\sim}29$~hr after discovery \citep{2023TNSTR1158....1I}. 

For each observation, raw images were dark-subtracted, if dark frames were taken, and plate-solved. Images that were off-target or had insufficient quality to plate solve were discarded. Calibrated images were aligned and averaged into stacked images with integration time $\approx60$~s. To separate the SN signal from the host galaxy, stacks were high-pass filtered using a median boxcar with width equal to 6 aperture radii. Differential aperture photometry was then performed on stacked images to measure fluxes of the SN and 3--5 reference stars with known {\it Gaia} magnitudes that were transformed to the Johnson-Cousins $V$ band (\citealt{2022arXiv220800211G}). The radius of the circular aperture was minimized to enclose $\geq90$\% of the reference stars' flux, which varied from $4\arcsec$ to $12\arcsec$ (3--7 pixels) to accommodate observing conditions.

The SN $m_V$ was then calculated as $m_V=m_{Vref} + 2.5\mathrm{log}(F_{ref})/F_{SN})$, where $m_{Vref}$ is the reference star's apparent $V$ magnitude and $F_{ref}$ and $F_{SN}$ are the measured reference star and SN fluxes, respectively. This was repeated for all reference stars and stacks. Measurements of the SN $m_V$ with signal-to-noise ratios $< 5$ were discarded, and we report the mean of the remaining magnitudes. The standard deviation of magnitudes within a given observation is the reported uncertainty. We consider observations where the SN $m_V$ exceeds the standard deviation of the background noise ($m_V - \sigma_{bg}$) to be non-detections.

\section{Results} \label{sec:data}

Figure \ref{fig:LC} contains 243 detections and 9 nondetection limits from 252 observations by 123 observers, 88 whom were identified because EV data are uploaded anonymously. Our earliest prediscovery detection was at 05-19 01:18, 4.83~hr after \citet{2023TNSAN.130....1M}'s detection and 16.15~hr before the discovery epoch. Additionally, \citet{2023TNSAN.123....1F} used a Unistellar telescope and were first to report a nonsurvey prediscovery detection at 2023-05-19 06:08.

We use a Markov Chain Monte Carlo (MCMC) sampling to obtain a best-fit model light curve with an exponential rise and linear decay. We obtain a peak $m_{v} = 11.05 \pm 0.08$ at 2023-05-25 21:37 $\pm$ 62~min, corresponding to $M_{V} = -18.18 \pm 0.09$~mag using a distance of $6.71 \pm 0.14$~Mpc and $E(B-V) = 0.031$~mag \citep{riess20162}. These values strongly agree with those found via meter-class telescopes (e.g., \citealt{Jacobson-Galan2023}). Our modeling implies an explosion time of 2023-05-18 22:43 $\pm$ 15~min, but this is 2.23~hr post earliest detection and not meaningfully constrained because our model is not well-fit to the initial rise (${\sim}1$ day, similar to \citealt{2023arXiv230606097H}).

\section{Conclusion} \label{sec:conc}

Here we present a light curve of SN 2023ixf with a 3.3~hr average sampling time over 35 days. Our modeled light-curve parameters support those gathered by professional telescopes and presented in other works. As such, this study provides crucial data to the community, but also demonstrates the power of a global observing network using telescopes with homogeneous opto-electronics, like Unistellar telescopes.

\vspace{5mm}
\facilities{Unistellar}
\vspace{15mm}


\bibliography{SN2023ixf}{}
\bibliographystyle{aasjournal}



\end{document}